\documentclass{evn2004}
\usepackage{txfonts} 

\begin{document}
   \title{EVN Observations of the BL\,Lac object ON~231
          \thanks{Also known as B1219+285} 
}
   \author{F.\,Mantovani\inst{1}
          \and
               E.\,Massaro\inst{2}
	  \and 
	       R.\,Fanti\inst{1}
	  \and
	       R.\,Nesci\inst{2}
	  \and
	       G.\,Tosti\inst{3}
	  \and
	       T.\,Venturi\inst{1}
}
   \institute{Istituto di Radioastronomia, CNR, 
              Via P. Gobetti, I-40129 Bologna, Italy 
         \and
Dipartimento di Fisica, Universit\`a di Roma ``La Sapienza", 
              Piazzale A. Moro 2, I-00185 Roma, Italy 
	 \and
	      Oss. Astronomico, Dipartimento di Fisica, Universit\`a di
	      Perugia, Via A. Pascoli, I-06123 Perugia, Italy
             }

   \abstract{
New EVN images at 5.0\,GHz and 8.4\,GHz of ON\,231 confirm the complex 
structure of the source, the identification of the core, 
the presence of components on the two sides of the core itself and a low 
brightness extension South-East of the main jet.
The optical behaviour of the source is discussed
in connection to changes in the radio structure.
   }

\maketitle
%

\section{Introduction}

BL Lacertae objects are characterized by large variations of their
radiation over a wide range of time scales from minutes to years.
The fast variability and the high apparent luminosity of these sources is
explained by Doppler boosting 
in a relativistic jet (Blandford and Rees 1978).
When historic light curves are available, it is found that
variations can also occur over time scales of several decades. 
One can expect, therefore, that the occurence of the largest 
and longest flares are associated with changes in the structure of the 
inner jet. To verify such a hypothesis we started a VLBI 
observation 
program of the bright BL Lac object ON\,231 (B1219+285; $z$=0.102) 
whose historic optical light curve was characterized by 
a long term brightening trend up to Spring 1998, when it reachead
a very high luminosity, followed by a dimming phase (Massaro et al. 1999, 
Tosti et al. 2002).
Massaro et al. (2001) presented VLBI images of ON\,231 obtained with 
the European VLBI Network (EVN) in February 1997 and June 1998, during 
the most recent active period. The source core was identified with the 
brigthest component because of its flat spectrum. According to this 
interpretation a new component, emerging from the core in the direction 
opposite to that of the main jet, was detected.
In this contribution we present new images at 5.0\,GHz and 8.4\,GHz of 
ON\,231. The new images confirm the unusual structure previously found
and possibly confirm the identification of the source core.


\section{Observations and Results}

The new VLBI images of ON\,231 were obtained during two
observing sessions scheduled by the European VLBI Network 
\footnote  {The European VLBI
Network is a joint facility of European, Chinese, South African and other
radio astronomy institutes funded by their national research councils.}
on 04 June 2001 at 5.0\,GHz and on 02 March 2002 at 8.4\,GHz. During the
8.4\,GHz session the geodetic stations of Matera and Wettzell joined the
network.
The raw data outputs from the correlators were calibrated in amplitude and
phase using AIPS and imaged using DIFMAP (Shepherd et al. 1995).
%
%
   \begin{figure}
   \centering
   \vspace{307pt}
   \includegraphics{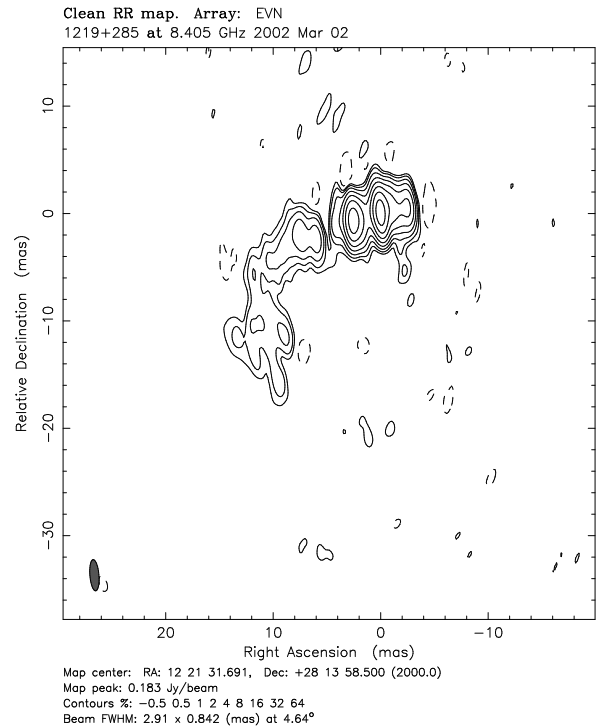}
   \caption{EVN image at 8.4\,GHz of ON\,231 (March 2002).}
         \label{fig:on2318}
    \end{figure}
\begin{table}[h]
\centerline{\bf Table\,1 - Modelfitting of the 5.0\,GHz and 8.4\,GHz images}
\vspace{0.1cm}
\hspace{0.0cm} 
\begin{tabular}{lrlrccr}
Comp.         & Flux~ & ~~R & PA~  & $a$  & Axial & $\phi$~~        \\
              & (mJy)& (mas)  & (deg)  & (mas)  & ratio & (deg)      \\
\hline
 2001  & June &    & (5\,GHz)   &    &    &     \\

 C0   & 210.6&  0.0  & 0.0    & 2.6   & 0.41  & 8.4    \\
 C1   &  88.6&  2.1  & 99.3   & 2.6   & 0.40  & 15.1   \\
 C2   &  48.7&  5.2  & 112.2  & 2.8   & 0.52  & 31.6   \\
 C3   &  32.8&  8.5  & 112.0  & 2.1   & 0.44  & 23.8   \\
 C4   &   5.2& 13.1  & 115.8  & 2.5   & 0.40  & 15.0   \\
 C5   &  78.7& 16.6  & 138.1  & 10.0  & 0.34  & 21.5   \\
 W1   & 142.2&  1.7  & $-$74.7 & 1.5  & 0.52  & 15.3   \\
 W2   &  29.1&  2.4  & $-$65.2 & 1.2  & 0.50  & 15.7   \\
 W3   &  20.1&  4.0  & $-$65.2 & 1.6  & 0.50  & 15.7   \\

\\

 2002  & March &    & (8.4 \,GHz) &    &    &     \\

 C0       & 268.2&  0.0  & 0.0    & 1.5  & 0.40  & 5.2  \\
 C1       & 136.7&  2.7  & 104.4  & 1.5  & 0.40  & 5.0  \\
 C2       &  51.3&  7.5  & 111.0  & 2.5  & 0.40  & 5.0  \\
 C3       &   5.3& 11.3  & 113.0  & 2.0  & 1.0   & 0.0  \\
 C4       &  41.7& 15.1  & 137.7  & 6.0  & 0.40  & 23.0 \\
 W1       &  53.8& 1.1   & $-$65.5 & 1.5 & 0.50  & 5.4   \\
 W2       &  30.1& 3.0   & $-$67.0 & 1.5 & 0.50  & 5.4   \\
\hline
\end{tabular}
\vspace{0.2cm}

$Note$: components C are ordered from West to East, components W from East
to West.
\end{table}
The new images show a source structure very similar to that described 
by Massaro et al. (2001): a jet elongating in the SE direction and 
bending toward South. The bright segment of the jet has a complex structure 
in which at least three components can easily be recognized with the 
brightest one in the middle (Fig.\,1). 
The main problem is the firm identification of the source core. 
We therefore fitted the jet structure of ON\,231 with gaussian components. 
For both epochs the modelling was limited to the 
brighter portion of the jet before the bending. Both results were well 
acceptable although a fraction of the diffuse emission along the jet could 
not be accounted for. The results of the modelfitting are presented in
Table\,1. The brightest component
is indicated by C0, those on the East side by C1, C2, ... while those on 
the West side by W1, W2, ...
From the data in Table 1 we can estimate the spectral indices of C0, C1
and of the entire West jet section. We prefer to consider the total
flux density from this section instead of those of the individual W components.
The resulting spectral indices ($S(\nu)\propto\nu^{\alpha}$)
are $\alpha_{C0}=0.46$, $\alpha_{C1}=0.82$, 
while the West component has the very steep $\alpha_W=-1.56$.
Fig. 2 shows the spectral index map obtained after the convolution of
the 8.4\,GHz image with the beam size achieved at 5.0\,GHz.  
The slightly inverted spectrum of C0 is evident and confirms what
measured in previous observations suggesting it is the core.
C1 also shows an inverted spectrum while previously showed a steep
spectrum. We suggest that, since the 8.4\,GHz
observations were done almost two years later than those at 5.0\,GH,
we are not comparing the flux densities from the same components 
(see Tab.\,1). 

%
   \begin{figure}
   \centering
   \vspace{307pt}
   \includegraphics{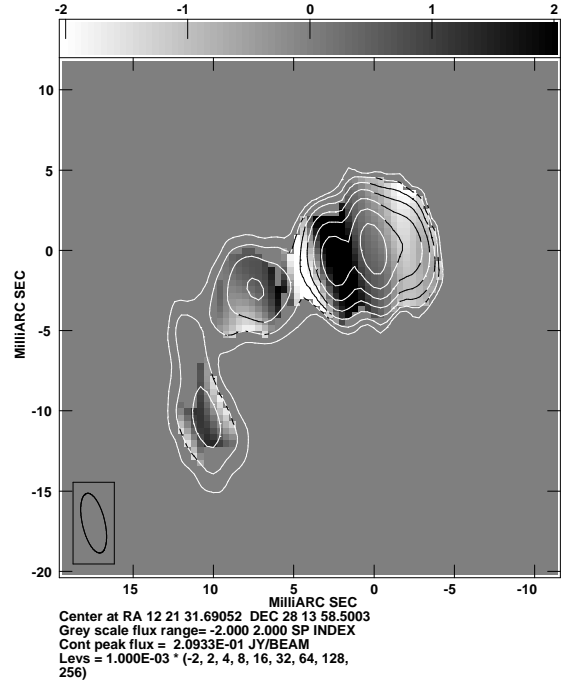}
   \caption{Grey scale map of the spectral index distribution of ON~231 
            derived from the two EVN images at 5.0\,GHz and at 8.4\,GHz 
            convolved with the same beam.
               }
               \label{fig:on2318}
   \end{figure}
\section{Discussion }
Parsec-scale jets of BL Lac objects are generally one-sided. The jet 
structures frequently show wiggles, suggesting a strong interaction
of the jet with the surrounding medium or motion like precession.
With the aim to correlate optical and radio behaviour for ON\,231,
the possible scenarios for interpreting both the time
evolution in the optical luminosity and the changes in the parsec scale 
structure are:

\smallskip\noindent
{$\bf i)$} 
the jet points very close to the observer direction, 
and undergoes strong instabilities and oscillations. The new western
component detected in ON\,231 may be therefore the result of one episodic 
large amplitude displacement, i.e. knots can be ejected from the nucleus 
along directions that are distributed around a mean position angle, so that 
one of these knots is displaced West of the core.

\smallskip\noindent
{$\bf ii)$} 
a slowly precessing jet, which approached the observer line of sight during 
the past few decades. The progressive decreasing of the angle would result in 
an increase in the beaming factor, and it could be responsible for the 
long brightnening trend detected in the optical. The minimum angular 
distance was likely reached in 1997-98, when ON 231 was observed at its maximum
brightness. 
This view is supported by a time scale analysis of the optical
variations, which reveals more rapid variability around that period
(Tosti et al. 2002). 
The jet direction 
changed to the opposite side of the core after this phase, as it appears in 
the EVN images.
The long, twisting extension of the jet after the bend at $\sim$12\,mas is
certainly not a young region of emission.
In the 
scenario of a precessing jet, it could be the trace of 
the jet itself, rotating clockwise from south to north.
\hfill\break\noindent

\begin{acknowledgements}
This research was supported by the European Commission's TMR Programme
``Access to Large-scale Facilities", under contract No.\ ERBFMGECT950012.
We acknowledge the support of the European Community - Access to Research
Infrastructure action of the Improving Human Potential Programme.
\end{acknowledgements}

\end{document}